\documentclass[twocolumn,pra,aps,superscriptaddress,english,floatfix]{revtex4-1}
\usepackage{amsmath}
\usepackage{amssymb}
\usepackage{amsfonts}
\usepackage{amssymb}
\usepackage[pdftex]{graphicx}
\usepackage{subfigure} 
\usepackage[english]{babel}
\usepackage{braket}
\usepackage{color}
\usepackage{mathtools}
\usepackage{threeparttable}
\usepackage[colorlinks,linkcolor=blue,anchorcolor=blue,citecolor=blue,urlcolor=blue]{hyperref}
\usepackage{babel}
\makeatother

\begin{document}

\title{Limits to single photon transduction by a single atom: Non-Markov
theory}

\author{Li-Ping Yang}

\affiliation{Birck Nanotechnology Center and Purdue Quantum Center,School of Electrical
and Computer Engineering, Purdue University, West Lafayette, IN 47906,
U.S.A.}

\author{Hong X. Tang}

\affiliation{Department of Electrical Engineering, Yale University, New Haven,
Connecticut 06511, USA}

\author{Zubin Jacob}
\email{zjacob@purdue.edu}

\homepage{http://www.zjresearchgroup.org/}


\affiliation{Birck Nanotechnology Center and Purdue Quantum Center,School of Electrical
and Computer Engineering, Purdue University, West Lafayette, IN 47906,
U.S.A.}

\begin{abstract}
Single atoms form a model system for understanding the limits of single photon detection. Here, we develop a non-Markov theory of single-photon absorption by a two-level atom to place limits on the absorption (transduction) time. We show the existence of a finite rise time in the probability of excitation of the atom during the absorption event which is infinitely fast in previous Markov theories. This rise time is governed by the bandwidth of the atom-field interaction spectrum and leads to a fundamental jitter in time-stamping the absorption event. Our theoretical framework captures both the weak and strong atom-field coupling regimes and sheds light on the spectral matching between the interaction bandwidth and single photon Fock state pulse spectrum.  Our work opens questions whether such jitter in the absorption event can be observed in a multi-mode realistic single photon detector. Finally, we also shed light on the fundamental differences between linear and nonlinear detector outputs for single photon Fock state vs. coherent state pulses. 

\end{abstract}
\maketitle

\section{Introduction}
Recently, single-photon detectors with high efficiency ($>90 \%$)~\cite{lita2008counting}, low dark count rate ($<1$ milli-Hz)~\cite{schuck2013waveguide}, and reduced timing jitter ($<20$ picoseconds)~\cite{pernice2012high} have raised substantial interest due to their widespread applications in quantum information processing~\cite{hadfield2009single,eisaman2011invited}, imaging, sensing/ranging and astronomy. Single-photon detectors based on superconducting nanowires or semiconductor avalanche photodiodes~\cite{liu2007high} are threshold detectors fundamentally different from conventional coherent photoreceivers. They typically work by outputting a classical electrical signal that results from the amplification of a weak quantum signal generated within the detector after the transduction event (photon absorption).

While the classical response of the amplifier stage currently dominates the characteristics of the single photon detector, it is important to note that the initial transduction process from photons to detector modes is fundamentally probabilistic in nature. As absorption mode volumes are decreased and amplifiers are improved, it is plausible that quantum limits of single photon transduction to detector modes will eventually manifest itself.  For example, currently timing jitter of a single photon detector is defined using the  deterministic output voltage signal where familiar concepts of rise time and fall time of classical electrical signals can be applied.  However, rising edges and fall times can also be defined for quantum signals during the initial single photon transduction event through the time-dependent excitation probability of detector modes in a statistical sense. Thus even an ideal single photon detector will be non-instantaneous and the rise times and falling edges of the quantum signal will set fundamental limits to the timing jitter performance. 

Glauber's theory of quantum photodetection and related work only provide the average counting rate of an ideal detector, which is proportional to the first order coherence of the electromagnetic field~\cite{glauber1963quantum,glauber2007quantum}.  In this class of approaches, the interaction between light and the ideal detector is based on first-order perturbation theory similar to Fermi's Golden rule. The properties of the detector, which is only weakly coupled to the electromagnetic field, thus cannot be characterized beyond averaged absorption and emission times (eg: Einstein's A and B coefficients for an atom in thermal equilibrium  ~\cite{loudon2000quantum}).  Thus this approach can not give any information on the rise time or fall time of the quantum or classical output signal of the detector. 

In this paper, we study the single atom as a model system to understand limits of the initial transduction event within an ideal narrowband single photon detector. Here, the output signal is the time-dependent excitation probability of the atom for an incident single photon pulse. We utilize an exactly solvable non-perturbative model to show the behavior of rise times and fall times of the output signal from such an atomic photodetector. We show that a non-Markovian theory is essential to understand the fundamental limit of the absorption (transduction) event. We also study the atom-field interaction beyond the weak coupling limit for various Fock state pulse envelopes and set limits on the rise times and fall times for atomic excitation probability. We expect these results to be a starting point for developing a complete theory for a multi-mode broadband single photon detector beyond weak interactions. 


\begin{figure*}
\includegraphics[width=16cm]{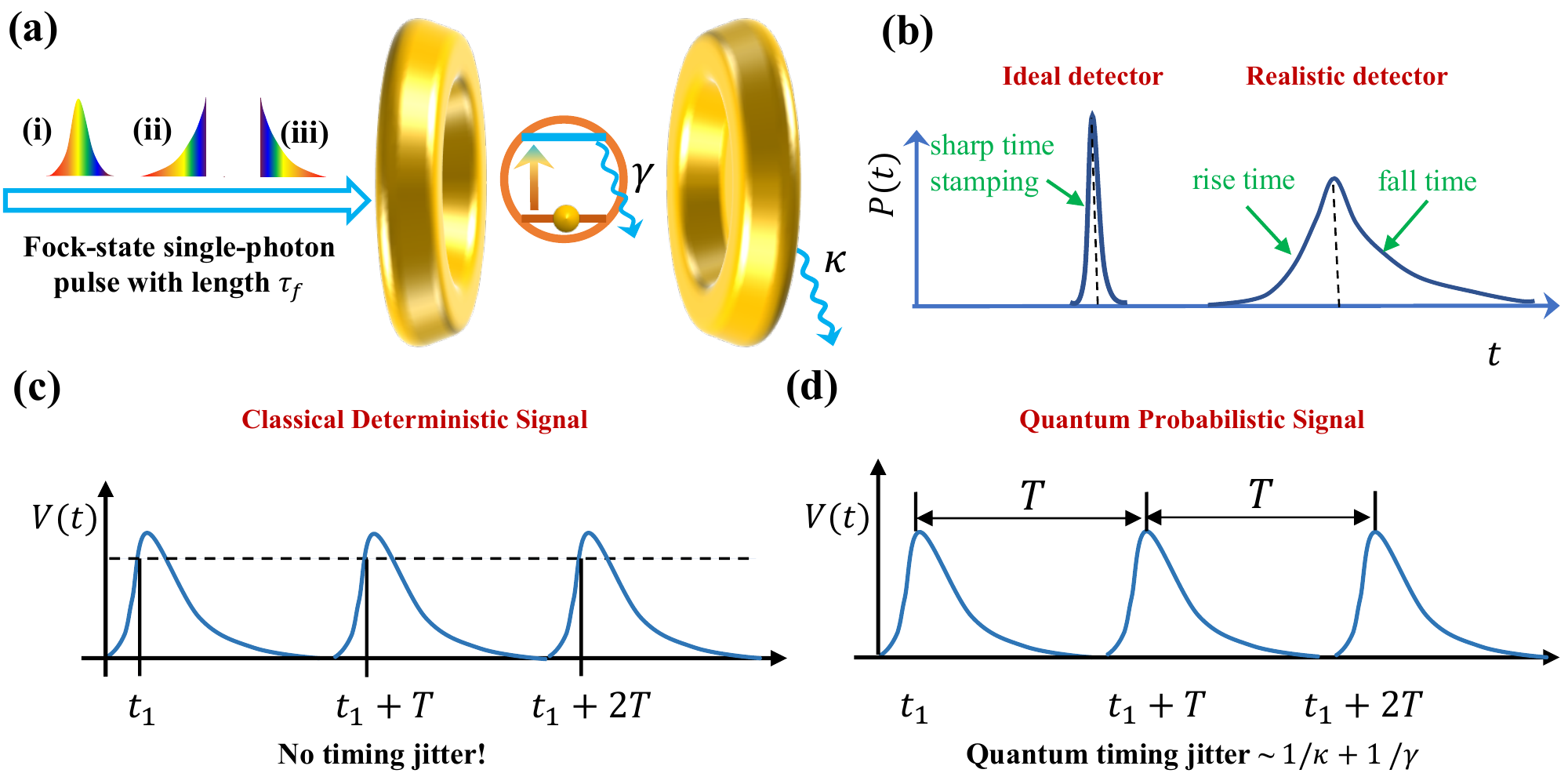}\caption{\label{fig:Schematic}(a) Schematic of a two-level atom, which functions as a narrow-band photo-detector, in a cavity excited by single-photon Fock-state pulses with envelope of Gaussian (i), exponential decaying (ii), and exponential rising (iii) shapes, respectively. 
Here, $\tau_f$ is the length of the propagating pulse, $\gamma$ the Markov spontaneous decay rate of the atom, and $\kappa$ the loss rate of the cavity. The atom-field interaction spectrum is modified by the cavity to be of Lorentzian form. (b) demonstrates the  quantum output signal $P(t)$ of an ideal photo-detector with infinite fast photon transduction and a realistic one, respectively. In contrast to the sharp time-stamping of a counting event in an ideal photo-detector, the finite rise/fall time of the quantum output signal leads to an intrinsic jitter of a realistic detector. Our theory places limits on the rise time which can be infinitely fast in Markovian approaches. In (c) and (d), we contrast between a classical deterministic signal sequence and a quantum probabilistic output sequence. For the classical deterministic signal (c) with perfect period $T$ and the same temporal profile, the clicks happen at the same position each time when the classical output pulses cross the threshold. Thus the click events are deterministic and no timing jitter exists. But for quantum signal sequence (d),  each pulse describes the excitation probability. The atom could be excited at any time during each pulse, which results in the quantum jitter.}
\end{figure*}

We note that the absorption~\cite{stobinska2009perfect,wang2011efficient,wang2012state,baragiola2012n} and scattering~\cite{domokos2002quantum,shen2005OptLett,zhou2008controllable,zhou2013quantum,longo2010few,shi2011two,zheng2010waveguide,rephaeli2012few,huang2013controlling,liao2015single,roulet2016solving,leong2016time}
of a propagating single-/multi-photon pulse by a single atom have
been extensively studied theoretically and also experimentally~\cite{tey2008strong,piro2011heralded,aljunid2013excitation,wenner2014catching,leong2016time}. In particular, important recent work has shown the role of multi-mode quantum pulses and interactions with atoms~\cite{wang2011efficient,baragiola2012n,domokos2002quantum}. However, in the case of ultrashort incident pulses, most previous theories are not directly applicable. This is due to the widely utilized Markov approximation which allows for instantaneous response times (or instantaneous detector response times) and thus cannot set limits to the single photon transduction event~\cite{stobinska2009perfect,wang2011efficient,wang2012state,baragiola2012n}. Note that the field of non-Markovian dynamics has recently given rise to widespread interest in the field of open quantum systems. In particular, non-Markovian theories have been previously utilized for single atoms in a cavity for memory applications and scattering in  one-dimensional (1D) wave-guides~\cite{shi2015multiphoton} as well as 1D lattices~\cite{ramos2016non}. However, the necessity of a non-Markov theory to explore the timing jitter of the single photon absorption event in an atom has not been realized.  We also discuss in our work the strong coupling regime of cavity QED which cannot be reached for a free space atom interacting with a single photon pulse. We also note that density of states engineering directly controls only the spontaneous emission i.e. fall time of the probability amplitude of the excited atom. Our focus in the paper is the fundamental limit in the absorption time or rise time in the excitation probability.

This paper is organized as follows. In Sec.~\ref{sec:SPATheory}, the model and non-Markov theory to describe single-photon transduction by a single atom is presented.  The limits to the transduction time and the concept of quantum timing jitter are given in Sec.~\ref{sec:TransductionLimit}. In Sec.~\ref{sec:SpectrumMatching}, we study the role of the spectral match between the Fock state pulse and the atom-field interaction spectrum. In Sec.~\ref{sec:nonlinear}, we study the fundamental difference between linear and nonlinear detector outputs for different types of quantum pulse. Finally, we collect our main conclusions in Sec.~\ref{sec:Conclusion}. The details of the wave function and 
spectral amplitude of a single-photon pulse is given in Appendix~\ref{sec:PhotonWaveFunction} 
and some detailed calculation about the excitation probability of the atom is presented in 
Appendix~\ref{sec:ExactSolution}.

\section{Non-Markov theory of single-photon absorption by a single atom\label{sec:SPATheory}}

In this section we define our model and provide the non-Markov theory of photon transduction before applying it to explore timing jitter. We study the absorption of a single photon Fock state (SPFS) pulse propagating in the $+z$ direction by a two-level
atom (playing the role of a nonlinear detector) at $z=z_{0}$. In
Coulomb gauge, the interaction between the pulse and the atom is given by~\cite{domokos2002quantum,wang2011efficient}
\begin{equation}
H_{{\rm a-p}}=i\hbar\int_{0}^{\infty}d\omega[g(\omega)e^{i\omega z_{0}/c}\hat{a}(\omega)\hat{\sigma}_{+}-{\rm h.c.}],\label{eq:H_int}
\end{equation}
where $\hat{a}(\omega)$ is the bosonic operator of the pulse mode
with frequency $\omega$, $\hat{\sigma}_{+}=|e\rangle\langle g|$
the Pauli operator of the atom with ground state $|g\rangle$, excited
state $|e\rangle$ and energy splitting $\omega_{d}$, and the rotating-wave
approximation has been taken. The amplitude of the atom-field interaction
spectrum is given by 
\begin{equation}
g(\omega)=\sqrt{\frac{\omega}{4\pi\varepsilon_{0}\hbar c\mathcal{A}}}(\vec{\epsilon}\cdot\vec{d}_{eg}),\label{eq:coupling_strength}
\end{equation}
where $\mathcal{A}$ is the effective transverse cross section of
the pulse~\cite{domokos2002quantum,blow1990continuum}, $\vec{d}_{eg}$
the electric dipole vector of the atom, and we have assumed that the
pulse is linearly polarized along the direction denoted by the unit vector
$\vec{\epsilon}$. Here, only the pulse modes have been taken into
account and the remainder field modes in the environment will be regarded
as a bosonic bath of the atom in the following. 

The interaction Hamiltonian between the atom and the bath modes is given by,
\begin{equation}
H_{{\rm a-b}}^{\prime}=i\hbar\int'd\omega[g'(\omega)e^{ik_{z}z_{0}}\hat{b}(\omega)\hat{\sigma}_{+}-{\rm h.c.}].\label{eq:H_int_prime}
\end{equation}
For a bath mode with wave vector $\mathbf{k}$ and frequency $\omega=c|\mathbf{k}|$,
the coupling coefficient $g'(\omega)$ has a similar form of $g(\omega)$
in Eq.~(\ref{eq:coupling_strength}). It should be pointed out that the prime of~$\int'd\omega$ means the the integral over $\omega$ includes the summation over different polarization and propagating directions of the bath modes with frequency
$\omega$~\cite{domokos2002quantum}, but excludes the pulse modes
which have already been taken into account in Eq.~(\ref{eq:H_int}).

In the case of SPFS pulse, the time evolution of the whole system can be solved
exactly for some special interaction spectra. The initial state of
the total system is assumed to be $|\psi(t_{0})\rangle=|g\rangle|1_{a}\rangle|0_{b}\rangle$,
where the atom is in the ground state $|g\rangle$, all the bath modes
are in the vacuum state $|0_{b}\rangle$, and the state of the pulse modes~$|1_{a}\rangle$ is given by~\cite{blow1990continuum},
\begin{equation}
|1_{a}\rangle=\int d\omega\xi(\omega)\hat{a}^{\dagger}(\omega)|0_{a}\rangle.\label{eq:SP_wavefunction}
\end{equation}
The normalization of the single-photon wave packet requires
\begin{equation}
\int d\omega|\xi(\omega)|^{2}=1.
\end{equation}
The details about the spectral amplitude $\xi(\omega)$ for SPFS
pulse of different envelope shapes is given in Appendix~\ref{sec:PhotonWaveFunction}.
The Hamiltonian~(\ref{eq:H_int}) and (\ref{eq:H_int_prime}) conserve
the total excitation number of the whole system, thus the wave function
of the total system at time $t$ can be expanded in the single-excitation
subspace as
\begin{eqnarray}
\!\!\!\!\!\!\!|\psi(t)\rangle & \!\!=\!\!\! & \left[\!\!\int\!\!\!d\omega A(\omega,t)\hat{a}^{\dagger}(\omega)\!+\!\!\!\int'\!\!\!\!\!d\omega B(\omega,t)\hat{b}^{\dagger}(\omega)\!\!+\!C(t)\hat{\sigma}_{+}\!\right]\!\!|G\rangle\!.\label{eq:TotalWaveFunction}
\end{eqnarray}
Here, $|G\rangle=|g\rangle|0_{a}\rangle|0_{b}\rangle$ is the ground
state of the whole system and the time-dependent coefficients $A(\omega,t)$,
$B(\omega,t)$, and $C(t)$ are determined by the Schr\"odinger equation
with initially conditions $A(\omega,t_{0})=\xi(\omega)$ and $B(\omega,t_{0})=C(t_{0})=0$.

After eliminating the degrees of freedom of the field, we obtain the equation
describing the exact dynamics of $C(t)$ (please refer to Appendix~\ref{sec:motion_eq}
for details),
\begin{eqnarray}
\frac{d}{dt}C(t) & = & -\int_{t_{0}}^{t}G(t-t')C(t')dt'\nonumber \\
 &  & +\int g(\omega)\xi(\omega)e^{i\omega z_{0}/c}e^{-i(\omega-\omega_{d})(t-t_{0})}d\omega,\label{eq:Ct}
\end{eqnarray}
where we have removed the fast oscillating factor $\exp(-i\omega_{d}t)$ of the probability amplitude $C(t)$ and used the initial condition $A(\omega,t_{0})=\xi(\omega)$. The
first and second terms on the r.h.s. of Eq.~(\ref{eq:Ct}) dominate
the decay and simulation behaviors of the atom, respectively. Similar non-Markov input-output theories have also be developed to handle single-photon storing in a cavity~\cite{shen2013single} and state transfer in a quantum network ~\cite{zhang2013non}.

All the memory effects on the decay behavior of the atom is contained in the integral
kernel~\cite{Yang2013master,cai2014threshold},
\begin{eqnarray}
G(t) & = & \!\!\!\int\!\!d\omega|g(\omega)|^{2}e^{i(\omega-\omega_{d})t}+\!\!\int'\!\!\!d\omega|g'(\omega)|^{2}e^{i(\omega-\omega_{d})t}\!.\label{eq:memory_kernel}
\end{eqnarray}
We first analyze the Markov approximation of previous works. The usual assumption entails that the band width of the interaction spectrum is much larger
than the coupling strength. Then the Wigner-Weisskopf
approximation~\cite{weisskopf1930berechnung,louisell1973quantum} applies,
which is proved to be equivalent to the Markov approximation~\cite{Yang2013master}.
By taking $|g(\omega)|^{2}$ and $|g'(\omega)|^{2}$ out of the integral, the kernel~$G(t)$ is reduced to a memory-less one~\cite{wang2011efficient,domokos2002quantum,gardiner1985input},
\begin{equation}
G(t-t')=(\gamma_{p}+\gamma')\delta(t-t'),
\end{equation}
where $\gamma_{p}=2\pi|g(\omega_{d})|^{2}$ and $\gamma'=2\pi|g'(\omega_{d})|^{2}$
characterize the decay of the atom back to pulse modes and the decay to bath
modes, respectively. Here we point to the delta function time-correlation which is the fundamental reason for instantaneous response times of the single atom. As shown in Appendix~\ref{sec:ExactSolution}, this approximation loses its validity for ultrashort pulses as well as the strong coupling regime ($\gamma\approx\kappa$) of cavity QED.

Instead, we consider here the atom to be
placed in a microcavity as shown in Fig.~\ref{fig:Schematic} (a). Usually, this increases the atom-field coupling strength. In
this case, the atom-field interaction spectrum is modified by the
cavity to be of Lorentzian form,
\begin{eqnarray}
|g_{{\rm tot}}(\omega)|^{2} & \!\!=\!\! & |g(\omega)|^{2}\!+\!|g'(\omega)|^{2}\!=\!\frac{\gamma/2\pi}{\left[(\omega-\omega_{d})/\kappa\right]^{2}\!+\!1},\label{eq:Lorentz_spectrum}
\end{eqnarray}
where the center of the interaction spectrum is assumed to be at $\omega_{d}$
for simplicity, the band width $\kappa$ is determined by the $Q$
factor of the cavity, and $\gamma=\gamma_{p}+\gamma'$ (the spontaneous decay rate under Markov approximation) characterizes
the atom-field coupling strength. In the bad-cavity limit, the incident pulse
shape will not change significantly by the cavity wall. Note here that we normalize the Lorentzian spectrum to compare with the Markov approximation (flat interaction spectrum) in the $\kappa \to \infty$ limit. This corresponds to the bad-cavity or weak-coupling limit of our theory.

After a somewhat
lengthy calculation, we finally arrive at a simple and intuitive exact
solution for $C(t)$ (the detailed calculation is shown in Appendix~\ref{sec:Exact_solution}), \begin{widetext}
\begin{eqnarray}
C(t) & = & s_{1}\left\{ e^{-p_{1}(t-t_{0})}C(t_{0})+e^{-p_{1}(t-t_0)}\int_{0}^{t-t_0}e^{p_{1}t'}\left[\int d\omega g(\omega)\xi(\omega)e^{i\omega z_{0}/c}e^{-i(\omega-\omega_{d})t'}\right]dt'\right\} \nonumber \\
 &  & +s_{2}\left\{ e^{-p_{2}(t-t_{0})}C(t_{0})+e^{-p_{2}(t-t_0)}\int_{0}^{t-t_0}e^{p_{2}t'}\left[\int d\omega g(\omega)\xi(\omega)e^{i\omega z_{0}/c}e^{-i(\omega-\omega_{d})t'}\right]dt'\right\} ,\label{eq:solution_C_t}
\end{eqnarray}
\end{widetext}where the decay rates~$p_j$ (also the frequency shift for
strong coupling case) and the ratios coefficients~$s_j$ for the two branches are given by Eqs.~(\ref{eq:pj}) and (\ref{eq:sj}), respectively. The $C(t_0)$ terms have been kept to study the spontaneous decay behavior of the atom.

Following from Eq.~(\ref{eq:solution_C_t}), we can see that the
memory kernel $G(t)$ [$p_j$ is determined by $G(t)$] not only determines the spontaneous decay behavior
of the atom, but also affects the fall time of the excitation probability of the atom $P(t)=|C(t)|^2$ in the photon
absorption process. This can be seen more clearly in the following. 

\begin{figure}
\includegraphics[width=8.5cm]{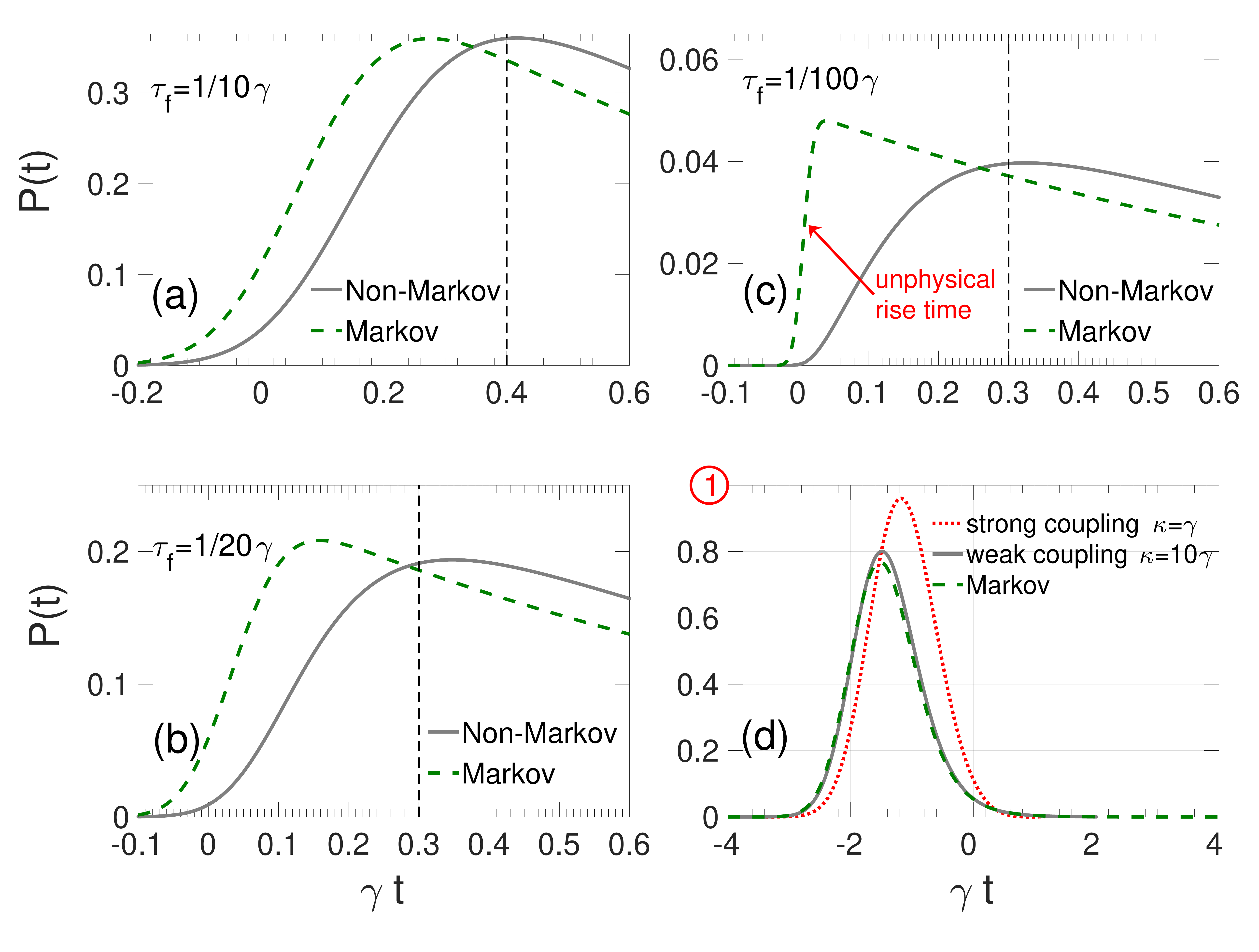}\caption{\label{fig2}The excitation probability~$P(t)$
of the atom stimulated by single-photon Fock-state pulse. Here, we take the Markov spontaneous 
decay rate of the atom as the unit~$\gamma=1$. (a-c) Comparison of $P(t)$ obtained by Markov theory (dashed-green lines) and non-Markov theory (solid-gray lines) for different pulse length. Here, we take the width of the interaction spectrum $\kappa$ as $\kappa=10\gamma$ for non-Markov theory and $\kappa\rightarrow \infty$ for Markov theory. The pulse lengths in (a-c) are taken as $\tau_f=1/10\gamma$, $\tau_f=1/20\gamma$, and $\tau_f=1/100\gamma$, respectively. As indicated by the red arrow in (c), the rise time of $P(t)$ under Markov approximation will go to zero when the pulse length~$\tau_f\rightarrow 0$. The solid-gray lines of the non-Markov clearly show there exists a finite rise time to the excitation probability of the single atom. (d) Comparison between the weak coupling and strong coupling regime of cavity QED. Here, the pulse length is set as $\tau_f=1/\gamma$. The dashed green line is obtained by the Markov theory. The solid gray line and dotted red line correspond to weak-coupling case ($\kappa=10\gamma$) and strong-coupling case ($\kappa=\gamma$), respectively. High excitation probability $P(t)> 0.96$ is obtained in the strong-coupling regime for the Gaussian pulse.} 
\end{figure}

\section{Limits to single-photon transduction by an atom \label{sec:TransductionLimit}}
Unlike the ideal photo-detector as shown in Fig.~\ref{fig:Schematic} (b), the output signal of a realistic detector always has a rise/fall time. In this section, we study the details of photon transduction by a single two-level atom. In experiments, the most commonly used waveform is a Gaussian wavepacket. In Fig.~\ref{fig2}, we present the time-dependent excitation probability $P(t)$ of the atom stimulated by a Gaussian
SPFS pulse. Fig.~\ref{fig2} (a-c) contrast the results obtained by the Markov theory (green lines) and our non-Markov theory (gray lines) for the weak coupling regime $\kappa=10\gamma$. We make the key observation that as the input pulse length ($\tau_f$) is decreased from $\tau_f=1/10\gamma$ to $\tau_f=1/100\gamma$, the excitation probability $P(t)$ obtained from the Markov theory has a rise time which directly follows the input pulse. This implies ultrashort pulses can cause sharp rise times in the single atom with no limit albeit with a low maximum in the excitation probability [see red arrow in Fig.~\ref{fig2} (c)]. This is a consequence of the Markov approximation by replacing~$g(\omega)$ with a constant~$\sqrt{\gamma_{p}/2\pi}$ in Eq.~(\ref{eq:solution_C_t})~\cite{domokos2002quantum,wang2011efficient}. A closer look made possible by our non-Markov theory shows that the rise time (gray lines) does not reduce as the pulse length is decreased. A clear bound exists on the fastest possible rise time even for a single atom excitation probability  $P(t)$. This rise time is limited by the atom-field interaction spectrum width~$\kappa$.

In Fig.~\ref{fig2} (d), we study the influence of the coupling strength on the atom excitation probability $P(t)$. In case of~$\tau_f\sim1/\gamma$, just as expected, there is no significant difference in the value of the probability $P(t)$ with (dashed green line)  or without (solid gray line) the Markov approximation in the weak-coupling case, while a significant variation occurs in the strong-coupling case (the dotted red line). The strong coupling enhances the excitation probability of the atom~$P(t)$ when~$\tau_f\sim1/\gamma$ for the Gaussian SPFS pulse.

\subsection{Concept of quantum jitter time for a single atom}

We elucidate here the concept of a quantum jitter time for a single atom. For the deterministic output signal of a classical system , eg: voltage signal of an RC circuit as shown in Fig.~\ref{fig:Schematic} (c), the characteristic rise time and fall time do not imply a timing jitter unless noise is added. In the case of a quantum output as shown in Fig.~\ref{fig:Schematic} (d), the probabilistic nature of the waveform implies there exists fundamental uncertainty in time-stamping an event such as photon transduction. For sake of discussion, we analyze the probability of excitation of the single atom as an output waveform. Under the previously used Markov approximation, we immediately notice that as the pulse length gets shorter, the rise time of the excitation probability is infinitesimal. This  implies, in principle, time-stamping of transduction is deterministic for a single photon by a single atom. However, our non-Markov theory predicts that even for a delta function input Fock state pulse, the rise time is finite and limited by the atom-field interaction spectrum bandwidth. Note the probability of excitation tends to zero for this narrow-band atomic detector but the rising rate (inverse of the rise time) is finite, not infinite. We emphasize that the finite rise time to the output waveform immediately implies there exists a quantum jitter in the absorption event. Thus even if the atom were completely characterized, the photon absorption event can occur anywhere within this finite rise-time and fall-time in subsequent experiments. 

In most systems, the weak coupling condition $\gamma\ll\kappa$ is well
satisfied. In this case, the Wigner-Weisskopf
approximation is applicable to the first the terms on the r.h.s. of Eq.~(\ref{eq:Ct})
by replacing the memory kernel $G(t-t')$ with $\gamma\delta(t-t')$.
But the same approximation can not be made on the second term, especially
when the width of the pulse spectrum~$1/\tau_{f}$ is of the same order of or even larger
than the interaction spectrum width $\kappa$. Then, the solution
for $C(t)$ is given by,
\begin{eqnarray}
\!\!\!\!\!\!\!\!C(t)\! & = & \!e^{-\frac{\gamma}{2}(t-t_{0})}\!\!\!\int_{0}^{t-t_{0}}\!\!\!\!\!\!\!\!\!\!dt'e^{\frac{\gamma}{2}t'}\!\!\!\!\int_{0}^{\infty}\!\!\!\!\!\!\!d\omega g(\omega)\xi(\omega)e^{i\frac{\omega z_{0}}{c}-i(\omega-\omega_{d})t'}\!\!.\label{eq:Ct_Markov}
\end{eqnarray}

\begin{figure}
\includegraphics[width=8cm]{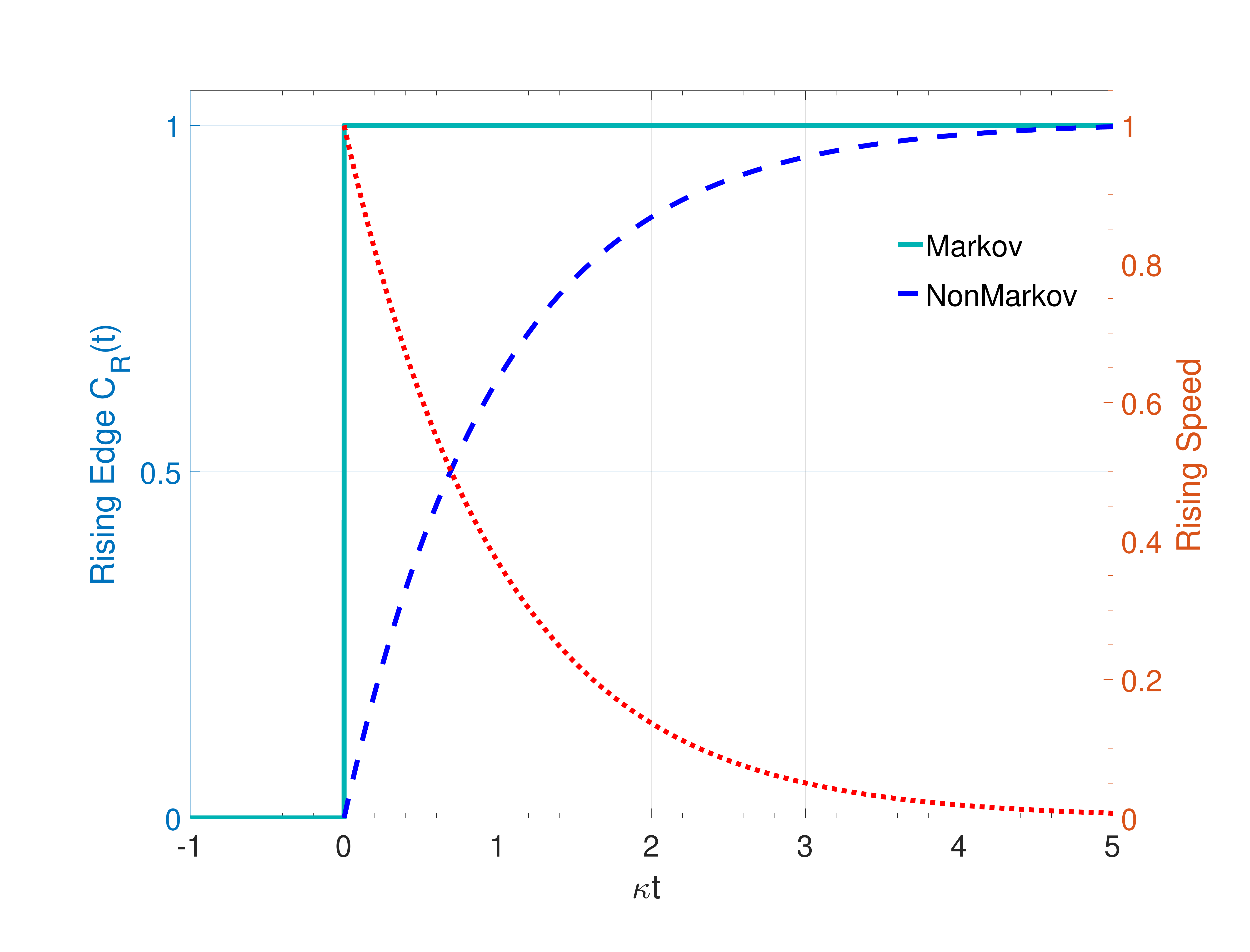}\caption{\label{fig3} For a~$\delta$-incident pulse, the rising edge~$C_{R}(t)$ of the probability amplitude with or without Markov approximation are displayed by the solid-light-blue line and the the dashed-blue line, respectively. The rising edge of the Markov~$C_{R}(t)$ is a step function with infinite large rising speed, while the rising edge of the non-Markov one has a finite increasing window. The corresponding finite rising speed $dC_{R}(t)/dt$ as depicted by the dotted-red line is limited by the finite interaction spectrum width~$\kappa$. This finite increasing speed with finite width places a limit on the rise time of the atomic photodetector. Note that for probabilistic quantum waveforms, this finite rise time implies a fundamental jitter different from deterministic classical output signals. }
\end{figure}

In order to get the analytic limit on photon transduction time, we take the Lorentzian interaction spectrum (\ref{eq:g_omega}) as an illustration and study a special case, where the length of the pulse is infinite short ($\delta$-pulse). In the case of weak-coupling~$\gamma_{p}\leq\gamma\ll\kappa$ using $\delta$-pulse with a constant spectrum amplitude $\xi(\omega)=\xi_{0}$, the solution of $C(t)$ reduces to a much simpler form 
\begin{equation}
C(t)=\xi_{0}\sqrt{2\gamma_{p}}e^{-\gamma (t-t_{0})/2}C_{R}(t),
\end{equation}
where the fall time of $C(t)$ is determined by the spontaneous decay
rate $\gamma$ and the rise time is determined by the monotonically increasing function,
\begin{equation}
C_{R}=\int_{0}^{t-t_{0}}\Theta(t'-z_0/c)\kappa e^{-\kappa(t'-z_{0}/c)}e^{\gamma t'/2}dt',
\end{equation}
with the Heaviside step function $\Theta(t)$. Note that in the Markov limit $\kappa\rightarrow\infty$, the function $\kappa\exp[-\kappa|t'-z_{0}/c|]\rightarrow\delta(t'-z_{0}/c)$,
then $C(t)$'s rising edge disappears (see the solid-light-blue line in Fig.~\ref{fig3}) and $C(t)$ decays instantaneously
after the arrival of the pulse (zero rise time), i.e., $C(t)=\xi_{0}\sqrt{\gamma_{p}}\exp[-\gamma(t-t_{0}-z_{0}/c)/2]$.
However, for a finite atom-field interaction spectrum $\kappa$, the
width of the rising window of $C_{R}(t)$ is determined by the width of its increasing speed, i.e., its time derivative 
\begin{equation}
\frac{d}{dt}C_{R}(t)=\Theta(t)\kappa e^{-\kappa t+\gamma t/2},
\end{equation}
where we have set $z_{0}=0$ for simplicity. In the weak coupling
limit $\gamma\ll\kappa$, the width of the rising speed $dC_{R}(t)/dt$
is approximated as $1/\kappa$ (see the dotted-red line Fig.~\ref{fig3}). Thus the width of the rising edge
of $C(t)$ as shown by the dashed-blue line in Fig.~\ref{fig3} as well as the the rising edge of the excitation probability
$P(t)$, has a lower limit $\sim 1/\kappa$.

\begin{figure*}
\includegraphics[width=16cm]{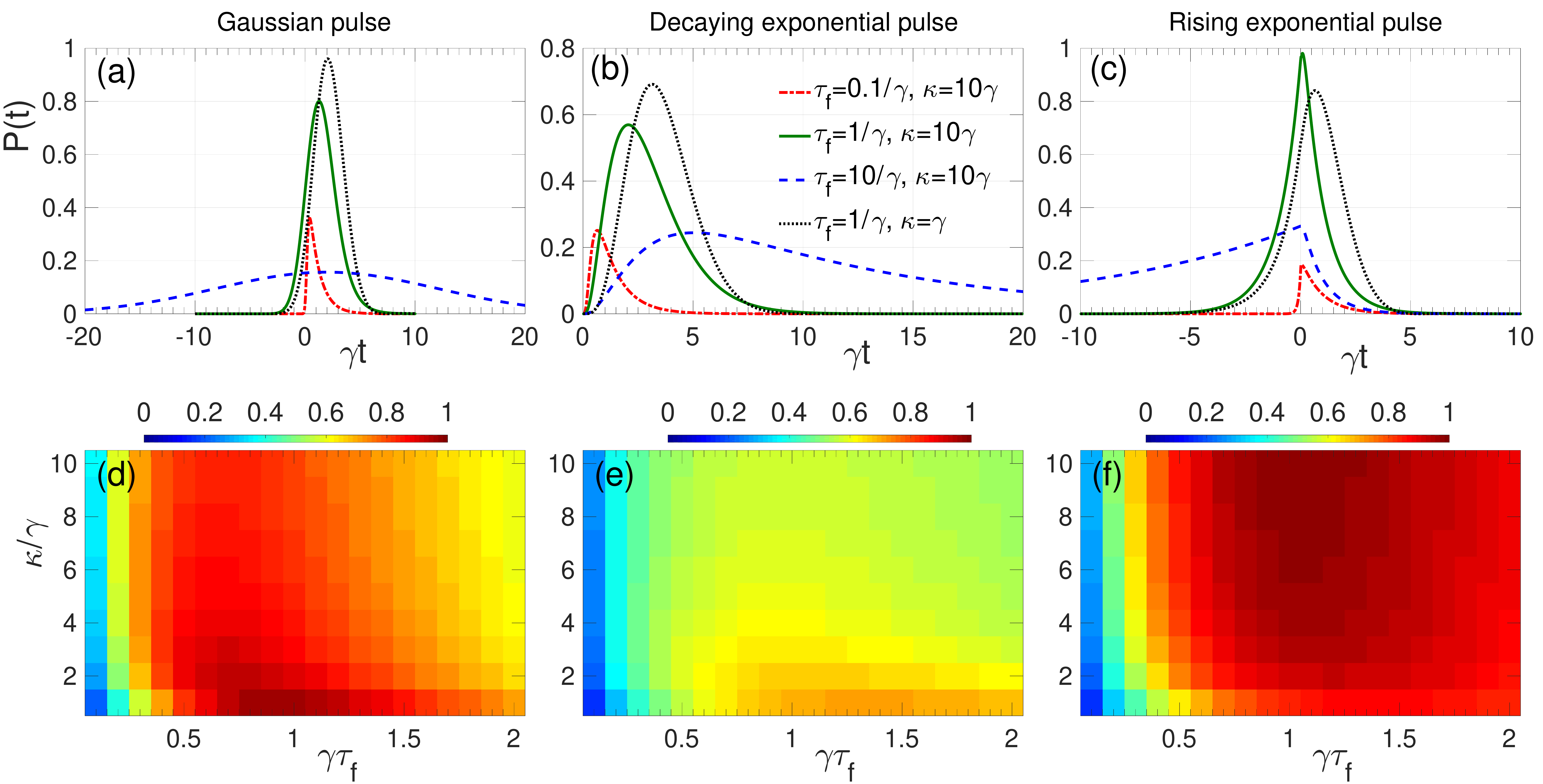}\caption{\label{fig4}The numerical results for Gaussian, decaying exponential, and rising exponential pulse are shown in the first to the third columns, respectively. In (a-c), the time-dependent excitation probability~$P(t)$ of the atom under different parameters is displayed. The parameters for (a-c) are the same and are shown in the legend of (b). In (d-f), the maximum of~$P(t)$ as a function of pulse length~$\tau_f$ and interaction spectrum width~$\kappa$ is presented. For a given $\kappa$, one can optimize the pulse length to get higher excitation probability and the global maximum is always located at $\tau_f\approx1/\gamma$ as indicated by Eq.~(\ref{eq:spectra_match}). }
\end{figure*}

We define the excitation probability density of the atom by normalizing $P(t)$,
\begin{equation}
\mathcal{P}(t)=\frac{P(t)}{\int P(t) dt}.
\end{equation}
The transduction time of a photon by an atom is characterized by the
width of envelope of the excitation probability density $\mathcal{P}(t)$, which equals to the sum of the widths of the rising and falling edges of $P(t)$. For weak coupling case, the fastest rise time and fall time of the atomic probability of excitation is given by
\begin{equation}
\Delta=1/\kappa+1/\gamma,
\end{equation} 
We note the first time constant corresponds to a finite rise-time (transduction) and the second constant corresponds to a fall time (spontaneous decay of the atom into field modes and bath modes).  For a realistic detector, the decay after the transduction is irreversible and will occur into the large set of amplifier modes. The fall time can also be directly controlled via density of states engineering. But the rise time is set by the bandwidth of detector-field coupling, which can not be eliminated. We can now analyze the consequence of the above equation for a solid state system. Note that the transduction event of light in a solid is governed by the electron-electron interaction which sets an upper limit (tens of femtoseconds for normal metals and semiconductors) for the rate of absorption \cite{guo2016universal,raman2013upper}. This will inevitably cause a jitter at least of the order of electron-electron interaction time.

\section{Spectral matching in photon transduction\label{sec:SpectrumMatching}}

For a single photon detector, fundamental characteristics such as quantum efficiency are considered independent of the input field.  This is evident in approaches such as detector tomography~\cite{lundeen2009tomography}, which are agnostic to the detector modes. Our concept of quantum timing jitter and spectral matching can be incorporated the formal theory of positive-operator valued measure (POVM)~\cite{van2017time,van2017photodetector}. As shown in the previous part, the excitation probability of detector modes and its time evolution in the transduction event depends on the spatio-temporal input Fock state and field-detector interaction spectrum. Thus it is probable that the spectral, spatial and temporal response characteristics of a single photon detector is sensitive to the chosen single photon waveform.  To explore this idea, we consider the role of the spectral overlap between the SPFS pulse and the atom-field interaction spectrum. From Eqs.~(\ref{eq:Ct}) and (\ref{eq:solution_C_t}),
we find that the temporal pattern of the rising edge of $C(t)$ is
determined by the overlap between the pulse spectrum $\xi(\omega)$ and the atom-field interaction spectrum $g(\omega)$. We show that the maximum of excitation probability
$P(t)$ can be optimized by matching these two spectra. Note we do not assume a flat/constant interaction spectrum as in previous works and our theory also works in the strong-coupling regime ($g>\gamma$ for a single photon).

The value of the excitation probability~$P(t)$ is determined by the overlap integral of interaction spectrum $g(\omega)$ and pulse spectrum $\xi(\omega)$, as can be seen from Eq.~(\ref{eq:solution_C_t}). In the frequency domain, the width of the pulse spectrum $|\xi(\omega)|^{2}$ is given by $1/\tau_{f}$. Thus, for a given interaction spectrum, the maximum excitation probability~$P(t)$ can be optimized by changing the pulse length. In Fig.~\ref{fig4}, we give the results of numerical optimization of $P(t)$. The three columns represent SPFS pulses with envelopes of Gaussian, decaying exponential, and rising exponential, respectively. In Fig.~\ref{fig4} (a-c), different lines represent different pulse length and different coupling strength (different interaction spectrum width~$\kappa$). Larger maximum of $P(t)$ can be obtained when the incident pulse has an optimized length~$\tau_{f}\approx1/\gamma$. For an optimized spectrally matched pulse~$\tau_{f}=1/\gamma$, the stronger coupling strength will enhance the maximum excitation probability $P(t)$ for Gaussian [Fig.~\ref{fig4} (a)] and decaying exponential pulses [Fig.~\ref{fig4} (b)]. However, this depresses $P(t)$ for rising exponential pulses [Fig.~\ref{fig4} (c)]. It should be pointed out that the maximum of $P(t)$ for Gaussian pulse obtained from our non-Markov theory $\sim0.96$ is larger than the Markov one $\sim0.8$ obtained in previous studies~\cite{wang2011efficient,baragiola2012n}.

In Fig.~\ref{fig4} (d-f), we plot the maximum of $P(t)$ as functions of pulse length~$\tau_{f}$ and interaction spectrum width~$\kappa$. We find that the global maximum of $P(t)$ is  always located at $\tau_{f}\approx1/\gamma$. To investigate the influence of the atom-field coupling strength on the global maximum of $P(t)$, we re-express the probability amplitude $C(t)$ in Eq.~(\ref{eq:solution_C_t}) as $C(t)=C_{1}(t)+C_{2}(t)$ with
\begin{equation}
C_{j}(t)=s_{j}\int d\omega \frac{g(\omega+\omega_{d})\xi(\omega+\omega_{d})}{p_{j}-i\omega}e^{-i\omega t}.\label{eq:spectra_match}
\end{equation}
Here, the integral variable $\omega$ is replaced with $\omega+\omega_{d}$,
the position of the atom is set as $z_{0}=0$ for simplicity, and
the lower limit of the integral over $t'$ has been extended to $-\infty$
without loss of generality. In the weak coupling case $\gamma\ll\kappa$,
one has $s_{1}\approx1,\ p_{1}\approx\gamma$ and $s_{2}\approx0,\ p_{2}\approx\kappa$.
Almost, only $C_{1}(t)$ has contribution to $C(t)$. In the strong
coupling region $\kappa\approx\gamma$, the decay rate of both $C_{1}(t)$
and $C_{2}(t)$ is given by $\Re[p_{j}]=\kappa\approx\gamma$. The
maximum of the probability amplitude $C(t)$ is determined by the
overlap of the interaction spectrum $g(\omega)$, the pulse spectrum
$\xi(\omega)$, and a factor $\sim1/(\gamma-i\omega)$. This is why the
global maximum of $P(t)$ is located at $\tau_{f}\approx1/\gamma$.

\begin{figure}
\includegraphics[width=8.5cm]{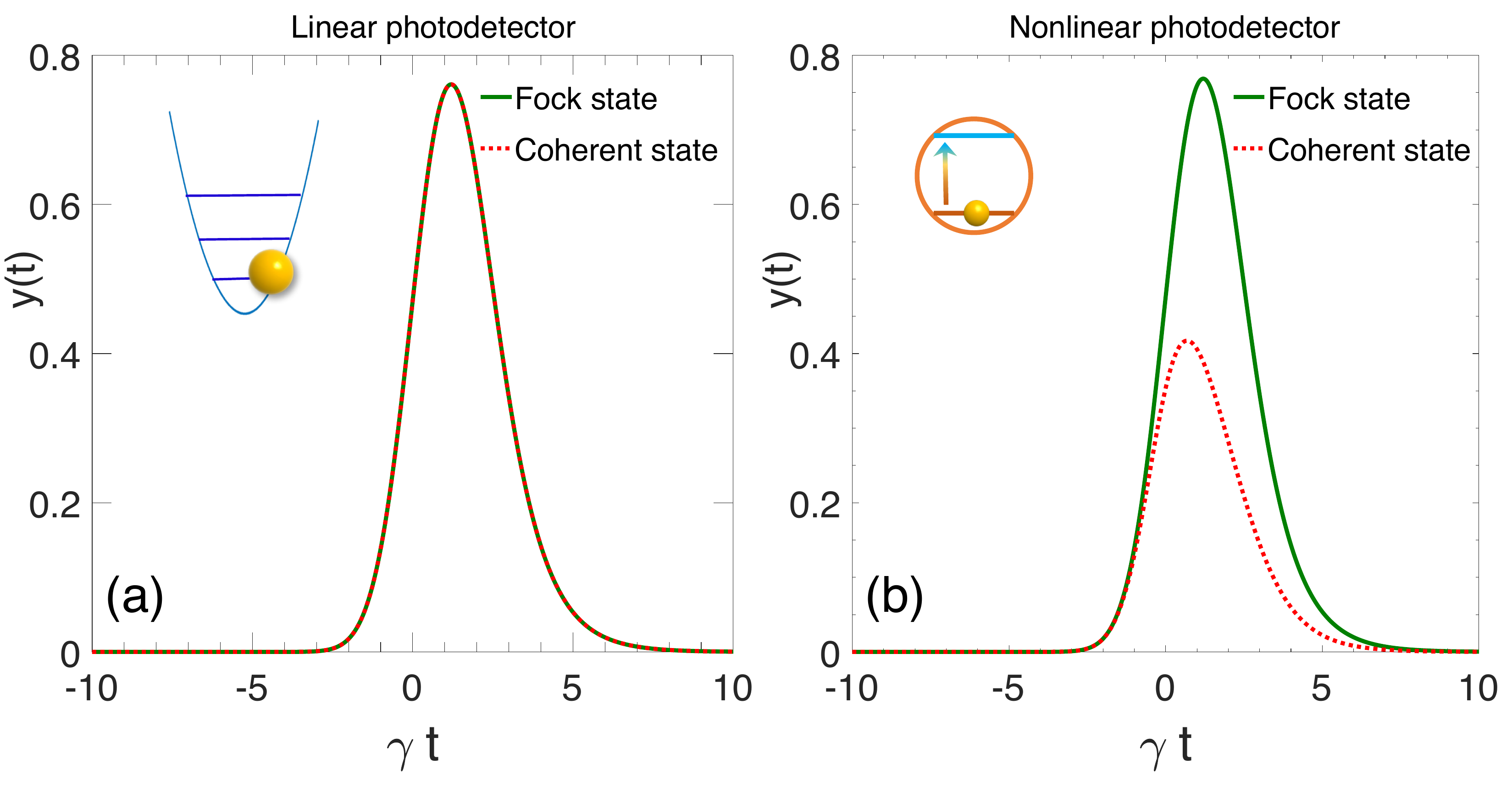}\caption{\label{fig5}The output waveforms $y(t)$ of linear (a) and nonlinear (b) photodetectors are displayed. The incident pulse is of Gaussian type with length~$\tau_f=1/\gamma$. The solid green line and dotted red line denote the single-photon Fock-state pulse and single-photon coherent-state pulse, respectively. All the results are obtained under the Markov approximation. The output waveform $y(t)$ of the linear photodetector for Fock-state puse and coherent pulse are the same. However, for the nonlinear photodetector,  the maximum of $y(t)$ for the coherent-state pulse is much smaller than that of the Fock state pulse. }
\end{figure}

\section{Linear photo-detector VS nonlinear detector \label{sec:nonlinear}}
In this section, we shed light on the difference between a linear photo-detector---a harmonic oscillator and a nonlinear one---single two-level atom. The Fock-state and coherent-state pulses have different statistical properties causing fundamental differences in excitation probabilities~\cite{loudon2000quantum}. The SPFS pulse has fixed photon number $\bar{n}=1$ and zero variance $\Delta n=0$. However, for a single-photon coherent-state (SPCS)  pulse with mean photon number $\bar{n}=1$, the corresponding variance is given by $\Delta n=\bar{n}$. As we show below, the harmonic-oscillator detector can not distinguish between the SPFS and SPCS pulse. Its output is linearly proportional to the strength of the incident pulse. But, for a single-atom detector, the  output waveforms of SPFS and SPCS pulses are very different.

The collective modes in a detector can be well modeled as harmonic oscillators. Here, we just consider one of the harmonic modes (described by the bosonic operators~$\hat{d}$ and~$\hat{d}^{\dagger}$), which is resonant with the center frequency of the incident pulse~$\omega_0$. The motion equation of this linear photo-detector is given in Appendix~\ref{sec:linear_detector}. The output waveform of this linear detector is given by the mean excitation in the detector $y(t)=\langle\hat{d}^{\dagger}(t)\hat{d}(t)\rangle$. As $y(t)$ is just proportional to the strength (mean photon number) of the incident pulse, the output waveform $y(t)$ of the linear detector are the same for SPFS and SPCS pulses as shown in Fig.~\ref{fig5} (a).

For the single-atom nonlinear detector, the output waveform is given by the excitation probability $y(t)=P(t)$, which has a upper limit $1$. Due to the conservation of excitation number, we can split the whole Hilbert space into many subspace. Each subspace has fixed excitation number and there is no cross talk between different subspaces. For a SPCS pulse, there is a large probability that no photon in this pulse (the zero-excitation subspace). And the multi-photon components in the pulse can  at most lead to one excitation in the detector, thus multi-photon components can not compensate for the loss resulting from the zero-photon component. Consequently, the excitation of the single-atom detector for SPCS pulse will be much lower than that for SPFS pulse as shown in Fig~\ref{fig5} (b). This result has also been found in Ref.~\cite{wang2011efficient}, however the contrast between linear and nonlinear outputs for different types of quantum pulses has not been shown till date and the underlying mechanism has not been revealed. It should be pointed out that all results in Fig~\ref{fig5} are obtained under the Markov approximation, since there is almost no difference between Markov and non-Markov ones in the case of optimal pulse length~$\tau_f=1/\gamma$ and weak coupling as shown in Fig.~\ref{fig2} (b).

\section{Conclusion \label{sec:Conclusion}}
In this paper, we have explored single photon transduction by a single 
atom using a non-Markovian theory to place bounds on the rise time of the atomic excitation probability. This upper limit to rise time is governed by the atom-field interaction spectrum and is not infinite as in previous theories. We note that the finite rise time and fall time of a probabilistic waveform implies a quantum jitter for atomic excitation probability. This arises even if the atomic excitation probability distribution is completely characterized experimentally since subsequent measurements of the absorption event can lie anywhere with the temporal width governed by the finite rise time and fall time. It is possible to scale up the model using dipole-dipole interactions between atoms \cite{cortes2017super} and also apply it to ultra-fast processes in quantum plasmonics \cite{jacob2012quantum}. In future work, we will address challenges of multi-mode and broadband single photon detectors building from insight of this narrow-band single-mode case. 

\begin{acknowledgments}
The authors thank Cristian L. Cortes, Joe Bardin, Josh Combes, and Ward Newman for fruitful discussions. This work is supported by DARPA DSO through a grant from ARO (W911NF-16-2-0151).  L.P.Y. has also been supported by China Postdoctoral Science Foundation (Grant No. 2015M580966).  
\end{acknowledgments}

\renewcommand{\baselinestretch}{1.0}
\makeatother

\appendix

\section{Wave function of the quantized pulse\label{sec:PhotonWaveFunction}}

The theory of quantization of the continuous-mode field is given by
Ref.~\cite{blow1990continuum}. In Coulomb gauge, the positive frequency
part of the electric field operator of the pulse can be expanded as~\cite{loudon2000quantum,blow1990continuum},
\begin{equation}
\hat{E}(z)=i\int_{0}^{\infty}d\omega\sqrt{\frac{\hbar\omega}{4\pi\varepsilon_{0}c\mathcal{A}}}\hat{a}(\omega)e^{i\omega z/c},\label{eq:A}
\end{equation}
where the constant $c$ is the speed of the light in the vacuum, $\varepsilon_{0}$
the permittivity of the vacuum, $\mathcal{A}$ the effective cross
section of the pulse, the field operators $\hat{a}(\omega)$ follow
the bosonic commutation relation, 
\begin{equation}
[\hat{a}(\omega),\hat{a}^{\dagger}(\omega')]=\delta(\omega-\omega').
\end{equation}
By taking the electric-dipole approximation, one can easily obtain
the interaction Hamiltonians between the atom and the continuous-mode
field in Eqs.~(\ref{eq:H_int}) and (\ref{eq:H_int_prime}).

The wave function of a single-photon Fock-state pulse with a spectral
amplitude $\xi(\omega)$ is given in Eq.~(\ref{eq:SP_wavefunction}).
The Fourier transform of the spectral amplitude $\xi(\omega)$ gives
the wave packet amplitude in the time-space domain,
\begin{equation}
\xi(t,z)=\frac{1}{\sqrt{2\pi}}\int d\omega\xi(\omega)e^{-i\omega(t-z/c)}.
\end{equation}
In this paper, the following four types of pulse are used:

(i) Gaussian pulse with wave packet amplitude in the time-space domain,
\begin{equation}
\xi(t,z)=\left(\frac{1}{2\pi\tau_{f}^{2}}\right)^{1/4}\exp\left[-\frac{(t-z/c)^{2}}{4\tau_{f}^{2}}-i\omega_{0}(t-z/c)\right],\label{eq:wavefunction_Gaussian}
\end{equation}
where $\tau_{f}$ is the pulse length and $\omega_{0}$ is the central
frequency, which is assumed to be resonant with the atom, i.e., $\omega_{0}=\omega_{d}$.
The corresponding spectral amplitude in the frequency domain reads,
\begin{equation}
\xi(\omega)=\left(\frac{2\tau_{f}^{2}}{\pi}\right)^{1/4}\exp\left[-\tau_{f}^{2}(\omega-\omega_{0})^{2}\right].
\end{equation}

(ii) Decaying exponential pulse with wave packet amplitude in the
time-space domain,
\begin{equation}
\xi(t,z)=\begin{cases}
0, & t<z/c\\
\sqrt{\frac{1}{\tau_{f}}}e^{-\frac{1}{2\tau_{f}}(1+2i\omega_{0}\tau_{f})(t-z/c)}, & t\geqslant z/c
\end{cases},
\end{equation}
and the corresponding spectral amplitude in the frequency domain,
\begin{equation}
\xi(\omega)=\sqrt{\frac{2\tau_{f}}{\pi}}\frac{1}{1-2i(\omega-\omega_{0})\tau_{f}}.
\end{equation}

(iii) Rising exponential pulse with wave packet amplitude in the time-space
domain,
\begin{equation}
\xi(t,z)=\begin{cases}
\sqrt{\frac{1}{\tau_{f}}}e^{\frac{1}{2\tau_{f}}(1-2i\omega_{0}\tau_{f})(t-z/c)}, & t\leqslant z/c\\
0, & t> z/c
\end{cases},
\end{equation}
and the corresponding spectral amplitude in the frequency domain,
\begin{equation}
\xi(\omega)=\sqrt{\frac{2\tau_{f}}{\pi}}\frac{1}{1+2i(\omega-\omega_{0})\tau_{f}}.
\end{equation}

(iv) Infinite short $\delta$-pulse, e.g., the pulse length of the
Gaussian type in Eq.~(\ref{eq:wavefunction_Gaussian}) goes to $0$.
The corresponding spectral in the frequency domain can be replaced
with a constant $\xi(\omega)=\xi_{0}\propto\tau_{f}$.

In addition, the wave function of a single-photon coherent-state pulse is given by
\begin{equation}
|\psi_{f}\rangle=\prod_{\omega}|\xi(\omega)\rangle,\label{eq:coh-state-pulse}
\end{equation}
where $|\xi(\omega)\rangle$ is the eigen state (coherent state)
of the field operator $\hat{a}(\omega)|\xi(\omega)\rangle =\xi(\omega)|\xi(\omega)\rangle$. The amplitude $\xi(\omega)$
satisfies 
\begin{equation}
\int d\omega|\xi(\omega)|^{2}=\bar{n}_f,
\end{equation}
with the mean photon number in each pulse $\bar{n}_f$ . 

\section{Exact dynamics of the photodetector\label{sec:ExactSolution}}

In this appendix, we give some detailed calculation of the dynamics
of the atom interacting with the continuous-mode field environment.
For the single-photon pulse with visible-wavelength, the thermal excitation can be neglected and the wave function of the total system
is expanded in the single-excitation subspace.

\begin{figure}
\includegraphics[width=8cm]{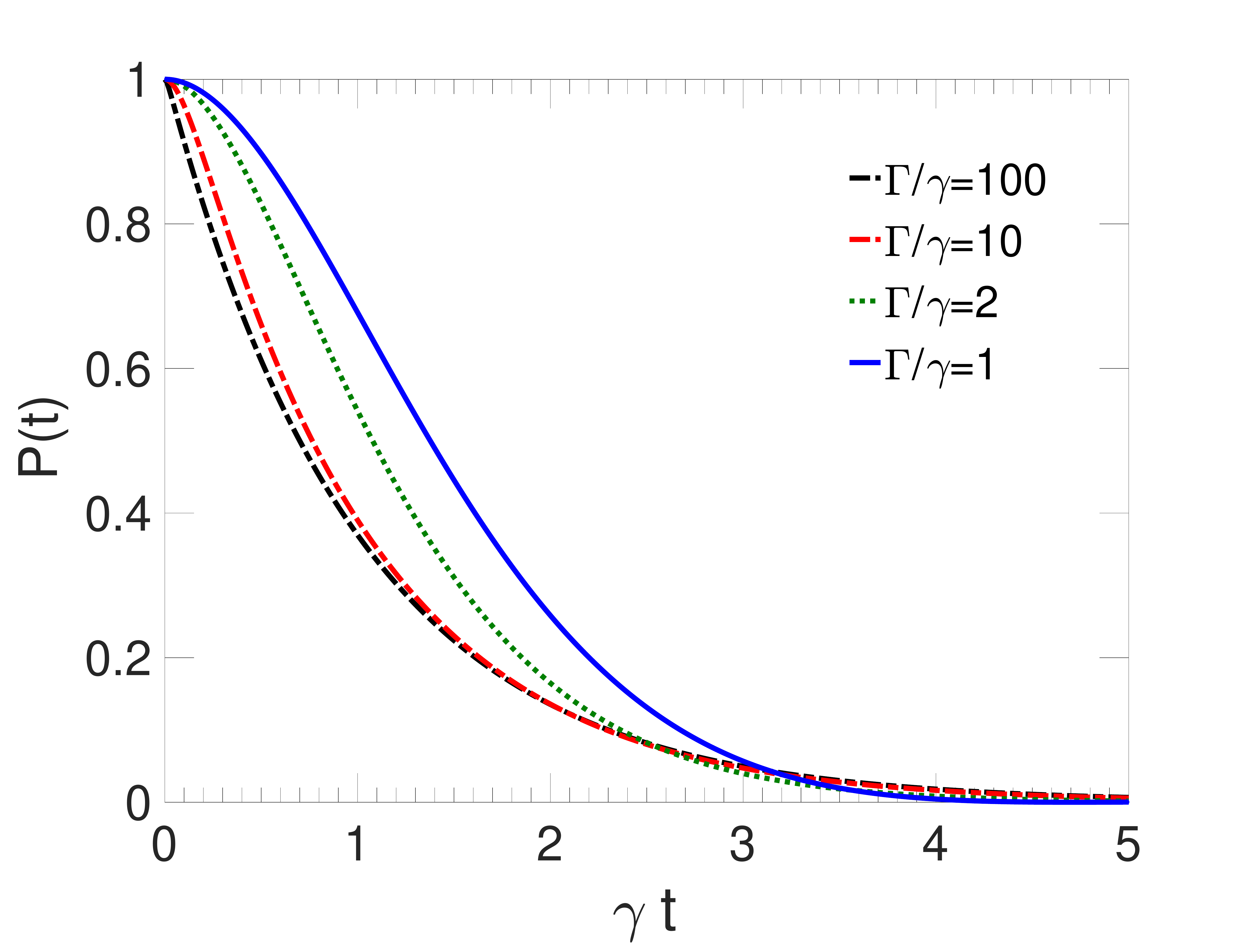}\caption{\label{fig6}The probability $P(t)$ of the atom in the excited state during its spontaneous decay process. Here, we take the Markov spontaneous
decay rate of the atom as the unit $\gamma=1.$ Different lines represent
different interaction spectrum width, from weak-coupling ($\kappa/\gamma=100$)
to strong coupling region ($\kappa/\gamma=1$).}
\end{figure}

\subsection{Motion equations\label{sec:motion_eq}}

The equations of the coefficients $A(\omega,t)$, $B(\omega,t)$,
and $C(t)$ in Eq.~(\ref{eq:TotalWaveFunction}) can be obtained
from the Schr\"odinger's equation as,
\begin{eqnarray}
\frac{d}{dt}A(\omega,t) & = & -g^{*}(\omega)e^{-i\omega z_{0}/c}C(t)e^{i(\omega-\omega_{d})(t-t_0)},\\
\frac{d}{dt}B(\omega;t) & = & -g^{\prime*}(\omega)e^{-ik_{z}z_{0}}C(t)e^{i(\omega-\omega_{d})(t-t_0)},\\
\frac{d}{dt}C(t) & = & \int d\omega g(\omega)e^{i\omega z_{0}/c}A(\omega,t)e^{-i(\omega-\omega_{d})(t-t_0)}\nonumber \\
 &  &\!\!\!\!\!\!\!\!\!\!\!\!+ \int'\!\!dg^{\prime*}(\omega)e^{ik_{z}z_{0}}B(\omega,t)e^{-i(\omega-\omega_{d})(t-t_0)}\!.\label{eq:original_Ct}
\end{eqnarray}
Here, these equations are obtained in the interaction picture, thus the fast oscillating factors in~$A$, $B$, and $C$ have been removed. The
formal solution of $A(\omega,t)$ and $B(\omega,t)$ reads
\begin{equation}
\!\!\!A(\omega,t)\! =  \!\xi(\omega)\!-\!\!\!\int_{t_{0}}^{t}\!\!\!dt'\!g^{*}(\omega)e^{-i\omega z_{0}/c}C(t')e^{i(\omega-\omega_{d})(t'-t_{0})}\!,
\end{equation}
\begin{equation}
\!\!B(\omega,t) =  -\int_{t_{0}}^{t}dt'g^{\prime*}(\omega)e^{ik_{z}z_{0}}C(t')e^{i(\omega-\omega_{d})(t'-t_{0})}, 
\end{equation}
where we have used the initial conditions $A(\omega,t_0)=\xi(\omega)$ and~$B(\omega,t_0)=0$. Substituting this formal solution into Eq.~(\ref{eq:original_Ct}),
the effective equation for $C(t)$ given in Eq.~(\ref{eq:Ct}) is
obtained. The total atom-field interaction spectrum is divided into
two parts $|g_{{\rm tot}}(\omega)|^{2}=|g(\omega)|^{2}\!+\!|g'(\omega)|^{2}$,
representing the coupling of the atom to the pulse modes and bath
modes respectively. Under the Wigner-Weisskopf approximation, these
two parts contribute two decay rates of $C(t)$~\cite{wang2011efficient},
\begin{eqnarray}
\gamma_{p} & = & 2\pi|g(\omega_{d})|^{2},
\end{eqnarray}
 and 
\begin{equation}
\gamma'=2\pi|g'(\omega_{d})|^{2}.
\end{equation}

In the following, we focus on the solution of $C(t)$ in the case
of Lorentzian interaction spectrum in Eq.~(\ref{eq:Lorentz_spectrum}).
The coupling amplitude $g(\omega)$ is chosen as
\begin{equation}
g(\omega)=\sqrt{\frac{\gamma_{p}}{2\pi}}\frac{1}{(\omega-\omega_{d})/\kappa+i}.\label{eq:g_omega}
\end{equation}
 For Lorentzian interaction spectrum, the corresponding memory kernel
$G(t-t')$ reads
\begin{eqnarray}
G(t) & = & \int d\omega|g_{{\rm tot}}(\omega)|^{2}e^{-i(\omega-\omega_{d})t}=\frac{\gamma\kappa}{2}e^{-\kappa|t|},\label{eq:G_t_1}
\end{eqnarray}
where $\gamma=\gamma_p+\gamma'$ and we have extend the low limit of $\omega$ in the integral (\ref{eq:G_t_1})
to $-\infty$, since $\kappa\ll\omega_{d}$. To get the results for strong atom-field coupling regime in the last
two sections of the paper, the interaction spectrum is taken to be of the Lorentzian
form. This limitation can be removed in the weak coupling case and
the excitation probability $P(t)$ can be easily obtained numerically for arbitrary
interaction spectrum $|g(\omega)|^{2}$.

\subsection{Exact Solution for Lorentzian spectrum\label{sec:Exact_solution}}

To get the solution of $C(t)$, we take the following Laplace transform
\begin{eqnarray}
f[p] & = & \mathcal{L}[f(t)]=\int_{0}^{\infty}e^{-pt}f(t)dt,\\
f(t) & = & \mathcal{L}^{-1}\left[f[p]\right]=\int_{\epsilon-i\infty}^{\epsilon+i\infty}e^{pt}f[p]dp,
\end{eqnarray}
of Eq.~(\ref{eq:Ct}). Then, we have
\begin{equation}
\!C[p]\!=\!F[p]\!\times\!\left[\!C(t_{0})+\!\!\int\!\!d\omega\frac{g(\omega)}{p+i(\omega-\omega_{d})}\xi(\omega)e^{i\omega z_{0}/c}\!\right]\!,
\end{equation}
where $C(t_{0})$ is kept to study the spontaneous decay behavior
of the atom and the singularities of 

\begin{eqnarray}
F[p] & = & \left[p+\frac{\gamma\kappa}{2}\frac{1}{p+\kappa}\right]^{-1},
\end{eqnarray}
will be used to carry out the inverse Laplace transform .

When $\kappa\neq2\gamma$, we can split $F[p]$ into
\begin{equation}
F[p]=\frac{s_{1}}{p+p_{1}}+\frac{s_{2}}{p+p_{2}}
\end{equation}
where 
\begin{eqnarray}
p_{j} & = & \frac{1}{2}\left[\kappa+(-1)^j\sqrt{\kappa^{2}-2\kappa\gamma}\right]\label{eq:pj},\ j=1,2
\end{eqnarray}
and the corresponding coefficients are
\begin{eqnarray}
s_{j} & = & \frac{1}{2}\left[1-(-1)^j\frac{1}{\sqrt{1-2\gamma/\kappa}}\right]\label{eq:sj}.
\end{eqnarray}
The real and imaginary parts of $p_{j}$ ($j=1,2$) correspond to
the decay rate and the Lamb shift, respectively. Then, the inverse
Laplace transform of $C[p]$ will gives the solution of $C(t)$ in
Eq.~(\ref{eq:solution_C_t}).

For the special case~$\kappa=2\gamma$, we have $p_{1}=p_{2}=\gamma$
and 
\begin{equation}
F[p]=\frac{1}{p+\gamma}+\frac{\gamma}{(p+\gamma)^{2}}.
\end{equation}
The inverse Laplace transform of $C[p]$ will gives\begin{widetext}
\begin{eqnarray}
C(t) & = & \left[e^{-\gamma(t-t_{0})}C(t_{0})+\int_{0}^{t-t_{0}}dt'e^{-\gamma(t-t')}\int d\omega g(\omega)\xi(\omega)e^{i\omega z_{0}/c}e^{-i(\omega-\omega_{d})t'}\right]\nonumber \\
 &  & +\left[\gamma(t-t_{0})e^{-p_{2}(t-t_{0})}C(t_{0})+\int_{0}^{t-t_{0}}dt'\gamma(t-t')e^{-\gamma(t-t')}\left[\int d\omega g(\omega)\xi(\omega)e^{i\omega z_{0}/c}e^{-i(\omega-\omega_{d})t'}\right]dt'\right].
\end{eqnarray}
\end{widetext}

In the weak-coupling case $\gamma\ll\kappa$, we can expand the
coefficients $p_{j}$ and $s_{j}$ to the first order of $\gamma/\kappa$,
\begin{eqnarray}
p_{1} & \approx & \frac{\gamma}{2},\ p_{2}\approx\kappa,\label{eq:weak-coupling_decay}
\end{eqnarray}
and
\begin{eqnarray}
s_{1} & \approx & 1+\frac{\gamma}{2\kappa},\ s_{2}\approx-\frac{\gamma}{2\kappa}.\label{eq:weak-coupling_coef}
\end{eqnarray}
The second branch has a much larger decay rate $p_{2}\gg p_{1}$ and
much smaller weight factor $s_{2}\ll s_{1}$. In this case, for $t>1/\kappa$,
we can neglect the contribution from the second branch. This is the
reason why a pure exponential spontaneous decay is observed in most
experiments.

To reveal the decay behavior of the atom, we study the spontaneous
process of the atom by resetting the initial conditions as $C(t_{0})=1$
and $A(\omega,t_{0})=B(\omega;t_{0})=0$. For~$\kappa\neq2\gamma$, the probability of the atom
staying in the excited state at time $t$ is given by
\begin{equation}
P(t)=|C(t)|^{2}=\left[s_{1}e^{-p_{1}(t-t_{0})}+s_{2}e^{-p_{2}(t-t_{0})}\right]^{2}.
\end{equation}
Usually, the excitation probability will not be a simple exponential decay. This non-exponential decay has already been found by Imamo\=glu
in 1994~\cite{imamog1994stochastic}. Only in the weak coupling case
$\gamma\ll\kappa$, the excitation probability approximately decay
as $P(t)\approx\exp[-\gamma(t-t_{0})]$ [see Eqs.~(\ref{eq:weak-coupling_decay}) and (\ref{eq:weak-coupling_coef})].
As shown in Fig.~\ref{fig6}, the probabilities of of the atom
in the excited state $P(t)$ from weak coupling (dash-dotted black
line) to strong (solid-blue line) coupling region are displayed. 

\subsection{Quantum Langevin equation for a linear detector\label{sec:linear_detector}}

In the previous part, the single-photon transduction by a nonlinear
detector (a two-level atom) have been studied. Similar method can be applied
to a linear detector case\textemdash a harmonic oscillator. If the
atom is replaced by a harmonic oscillator with frequency $\omega_{d}$,
we can easily obtain the quantum Langevin equation equation for the
annihilation operator $\hat{d}$ of the the oscillator\cite{gardiner1985input},
\begin{equation}
\frac{d}{dt}\hat{d}(t)=-\int_{t_{0}}^{t}G(t-t')\hat{d}(t')dt'+\hat{a}_{in}(t)+\hat{b}_{{\rm in}}(t),\label{eq:bt}
\end{equation}
where the memory kernel $G(t)$ is the same as the one in Eq.~(\ref{eq:memory_kernel})and the operators
\begin{equation}
\hat{a}_{in}(t)=\int d\omega g(\omega)\hat{a}(\omega,t_{0})e^{i\omega z_{0}/c}e^{-i(\omega-\omega_{d})t},
\end{equation}
and 
\begin{equation}
\hat{b}_{{\rm in}}(t)=\int'd\omega g'(\omega)\hat{b}(\omega,t_{0})e^{-ik_{z}z_{0}}e^{-i(\omega-\omega_{d})t},
\end{equation}
represent the input from the pulse modes and bath modes, respectively. As a generalization of the traditional input-output theory~\cite{gardiner1985input}, the signal and noise field modes are handled separately. Different from the noise operator~$\hat{b}_{in}(t)$, the input signal operator~$\hat{a}_{in}(t)$, which is determined by the incident pulse, can not be simply treated as a white noise operator.

Finally, it should be point out that, to get a larger excitation probability
$P(t)$ and to compare the results obtained by our non-Markov theory
with the previous Markov ones~\cite{wang2011efficient,baragiola2012n},
we let $\gamma_{p}=\gamma$ during the calculation. Thus, all results
obtained in this paper should be modified by a factor $\gamma_{p}/\gamma$.
This factor denotes the proportion of the pulse modes in the total
field modes. For three-dimensional free space, we assume that the
polarization of the pulse is parallel to the direction of the electric
dipole.  In this case, the factor reads $\gamma_{p}/\gamma=3/8\pi$.
For 1D wave guide, if the cross section the is small enough, the modes
with wave vector in the $xy$-plane $(\vec{k}\cdot\vec{z}=0)$ can
be neglected. In this case, the factor is $\gamma_{p}/\gamma=1/2$.
Usually, to realize an efficient coupling of photons to a single atom
in experiment, the incident light pulse is focused onto the atom by
a antennas~\cite{wrigge2008efficient,novotny2011antennas} or a parabolic
mirror~\cite{lindlein2007new}. The factor $\gamma_{p}/\gamma$ can
reach near $100\%$~\cite{sondermann2007design}.

\bibliography{PD}

\end{document}